\documentclass[12pt]{iopart}
\usepackage{graphicx}
\begin{document}
\title{Gravitational Collapse of Thick Domain Walls}
\author{David Garfinkle}
\address{Dept. of Physics, Oakland University,
Rochester, MI 48309, USA}
\address{and Michigan Center for Theoretical Physics, Randall Laboratory of Physics, University of Michigan, Ann Arbor, MI 48109-1120, USA}
\ead{garfinkl@oakland.edu}
\author{Ryan Zbikowski}
\address{Dept. of Physics, Oakland University,
Rochester, MI 48309, USA}
\ead{rmzbikow@oakland.edu}
\begin{abstract}
Numerical simulations are performed of the gravitational collapse of a scalar field with a $\lambda {\phi^4}$ potential. 
Comparisons are made with the thin shell approximation.
\end{abstract}
\maketitle
\section{Introduction}
\qquad
Domain walls in general relativity\cite{ipser} are usually treated in the thin shell approximation.\cite{israel}  
However, there is also a description
of a domain wall as a soliton of a field theory.  In a certain limit one expects the field theory description to reduce to the 
thin shell description.  Indeed, one can formally expand the Einstein-scalar equations in powers of the thickness and obtain 
the thin shell equations at lowest order in this expansion.\cite{ruthandme1,ruthandme2}  Nonetheless, one cannot simply 
assume the validity of such an expansion.  Instead in a field theory one gets to choose initial data and then the field
equations give the results of the evolution of those data.  Thus the thin shell approximation is a good approximation for a 
thick domain wall to the extent that the evolution of initial data that are well approximated by the thin shell treatment continue under evolution to be well approximated by this treatment.  To test this approximation, we will choose initial data 
for a spherically symmetric thick domain wall, perform a numerical evolution, and compare to the corresponding thin 
shell approximation.  Methods are described in section 2, results in section 3 and conclusions in section 4.  

\section{Methods}

The Lagrangian for a thick wall takes the form
\begin{equation}
L = - {\textstyle {\frac 1 2}} {\nabla ^a}\phi {\nabla _a} \phi - V(\phi)
\end{equation}
where $V(\phi)$ is a potential with two minima.  
The equation of motion associated with this Lagrangian is 
\begin{equation}
{\nabla ^a}{\nabla _a} \phi - {\frac {dV} {d\phi}} = 0
\label{wave}
\end{equation}
The shell also satisfies the Einstein field equation
\begin{equation}
{G_{ab}} = \kappa {T_{ab}}
\end{equation}
Here $G_{ab}$ is the Einstein tensor, $\kappa = 8 \pi G $ where $G$ is Newton's gravitational constant, and the stress-energy
of the scalar field is given by
\begin{equation}
{T_{ab}} = {\nabla _a}\phi {\nabla _b} \phi - {\textstyle {\frac 1 2}} {g_{ab}} ( {\nabla ^c}\phi {\nabla _c}\phi
+ 2 V )
\label{stressenergy}
\end{equation}
Simulations of such thick walls in the case where their self-gravity can be neglected were done 
by\cite{widrow} and \cite{spergel}.
We will use the standard $\lambda {\phi^4}$ potential
\begin{equation}
V = \lambda {{({\phi^2} - {\eta ^2})}^2}
\end{equation}

For simulations in spherical symmetry, one often takes the area radius as one of the spatial coordinates and chooses time
to be orthogonal to this radial coordinate.\cite{matt}  
However, this coordinate system breaks down when a trapped surface forms and
thus cannot follow the evolution after black hole formation.  Instead we will use the method of \cite{ratinryome} and 
use maximal slicing with radial length as the radial coordinate.  Maximal slicing allows us to simulate part of the region 
inside the black hole without encountering the singularity.  
The metric takes the form
\begin{equation}
d {s^2} = - {\alpha ^2} d{t^2} + {{(dr + {\beta ^r} dt)}^2} + {R^2} (d {\theta ^2} + {\sin ^2}\theta d {\varphi ^2})  
\label{metricform}
\end{equation}
Note that the usual area radius $R$ is not one of the coordinates and is instead a function
of the coordinates $t$ and $r$.  Thus the spatial metric $\gamma _{ab}$ has components
\begin{eqnarray}
{\gamma _{rr}}=1
\nonumber
\\
{\gamma _{\theta \theta }} = {R^2}
\nonumber
\\
{\gamma _{\varphi \varphi}} = {R^2} {\sin ^2} \theta
\end{eqnarray}
The extrinsic curvature, $K_{ab}$ is defined by 
\begin{equation}
{K_{ab}} = - {{\gamma _a}^c}{\nabla _c}{n_b}
\label{extrinsic}
\end{equation}
where $n^a$ is the unit normal to the surfaces of constant time $t$.  However, due to spherical symmetry and maximal
slicing, there is only one independent component of the extrinsic curvature.  Specifically we have
\begin{equation}
{{K^\theta}_\theta}={{K^\varphi}_\varphi} = - {\textstyle {\frac 1 2}} {{K^r}_r}
\end{equation}
Equation (\ref{extrinsic}) is equivalent to 
\begin{equation}
{\partial _t} {\gamma _{ij}} = - 2 \alpha {K_{ij}} + {D_i}{\beta _j} +{D_j}{\beta _i}
\label{gammadot}
\end{equation}
where $D_i$ is the covariant derivative of the spatial metric $\gamma _{ij}$.  The $rr$ component of 
eqn (\ref{gammadot}) yields
\begin{equation}
{\partial _r}{\beta ^r} = \alpha {{K^r}_r}
\end{equation}
whose solution is 
\begin{equation}
{\beta ^r} = {\int _0 ^r} \alpha {{K^r}_r} \; d r
\label{shift}
\end{equation}
The $\theta \theta $ component of eqn (\ref{gammadot}) yields
\begin{equation}
{\partial _t} R = {\beta ^r} {\partial _r} R + {\frac \alpha 2} R {{K^r}_r}
\label{dtR}
\end{equation}

We now use the momentum constraint of the Einstein field equation to determine the extrinsic curvature.  For maximal slicing
($K=0$) this constraint is
\begin{equation}
{D_a}{K^{ab}} = - \kappa  {\gamma ^{bc}}{n^d}{T_{cd}}
\label{momentum}
\end{equation}
Define the quantities $P$ and $S$ by 
\begin{equation}
P = {n^a}{\nabla _a} \phi , \; \; \; S = {\partial _r} \phi
\label{PSdef}
\end{equation}
Then eqn (\ref{momentum}) becomes 
\begin{equation}
{\partial _r} {{K^r}_r} + 3 {R^{-1}} {{K^r}_r} = - \kappa  P S
\label{Krr}
\end{equation}

Note that there is also a Hamiltonian constraint associated with the Einstein field equation.  In the case of maximal slicing, 
this constraint is 
\begin{equation}
{{}^{(3)}}R - {K_{ab}}{K^{ab}} = 2 \kappa {T_{ab}}{n^a}{n^b}
\end{equation} 
where ${{}^{(3)}}R$ is the spatial scalar curvature.  This equation yields
\begin{equation}
{\partial _r}{\partial _r} R = {\frac {1 - {{({\partial _r}R)}^2}} {2R}} 
- {\textstyle {\frac 1 4}} R \left [ 
{\textstyle {\frac 3 2}} {{({{K^r}_r})}^2} +  \kappa ({P^2} +{S^2}+ 2 V) \right ]
\label{hamilton}
\end{equation}

We now determine the lapse $\alpha$.  It follows from the maximal slicing condition that 
\begin{equation}
{D_a}{D^a} \alpha = \alpha \left [ {K_{ab}}{K^{ab}} + {\frac \kappa 2} {T_{ab}}({n^a}{n^b}+{\gamma ^{ab}})\right ]
\end{equation}
which yields
\begin{equation}
{\partial _r}{\partial _r} \alpha + {\frac 2 R} ({\partial _r}R )({\partial _r}\alpha )
= \alpha \left [ 
{\textstyle {\frac 3 2}} {{({{K^r}_r})}^2} + \kappa ({P^2} - V)\right ]
\label{lapse}
\end{equation}

We now consider the evolution of the scalar field.  From the definitions of $P$ and $S$ it follows that
\begin{eqnarray}
{\partial _t} \phi = \alpha P + {\beta ^r} S
\label{dtphi}
\\
{\partial _t} S = \alpha ({\partial _r}P + {{K^r}_r} S) + P {\partial _r} \alpha + {\beta ^r}{\partial _r} S
\label{dtS}
\end{eqnarray}
The equation of motion, eqn (\ref{wave}) becomes after some straightforward but tedious algebra
\begin{equation}
{\partial _t} P = {\beta ^r}{\partial _r} P + S {\partial _r} \alpha
+ \alpha \left [ {R^{-2}} {\partial _r} ({R^2} S) - {\frac {dV} {d\phi}}\right ]
\label{dtP}
\end{equation}

The initial data are as follows: we choose a moment of time symmetry so that $P$ and $K_{ab}$ vanish.  The scalar
field $\phi$ is chosen to have a flat spacetime domain wall profile with the center of the wall at a radius $r_0$.  More 
precisely, define the quantities $\epsilon$ and $\sigma$ by
\begin{eqnarray}
\epsilon &=& {\frac 1 {\eta {\sqrt {2\lambda}}}}\\
\sigma &=& {\textstyle {\frac 4 3}} {\sqrt {2\lambda}}{\eta ^3}
\end{eqnarray}
where $\eta$ and $\lambda $ are the parameters of the potential.  
Here $\sigma $ is the energy per unit area of the wall and $\epsilon$ is an effective wall thickness.  Formally the 
thin shell limit of the solution is the limit as $\epsilon \to 0$ at constant $\sigma$.  
Correspondingly, one can specify $\epsilon$
and $\sigma$; then $\lambda$ and $\eta$ are determined by
\begin{eqnarray}
\lambda &=& {\frac 2 {3 \sigma {\epsilon ^3}}}\\
\eta &=& {\sqrt {{\textstyle {\frac 3 4}} \sigma \epsilon}}
\end{eqnarray}
In flat spacetime a static planar domain wall 
solution is given by
\begin{equation}
\phi = \eta \tanh (z/\epsilon )
\end{equation}
We choose for initial data 
\begin{equation}
\phi = \eta \tanh ((r-{r_0})/\epsilon)
\end{equation}
where the constant $r_0$ can be chosen arbitrarily but should be chosen to be much larger than $\epsilon$.
The quantity $S$ is set equal to ${\partial _r} \phi$.  
Eqn. (\ref{hamilton}) is integrated for $R$ using the fact that at the origin $R=0$ and ${\partial _r}R=1$.  
Then at each time step, the evolution proceeds as follows: first eqns (\ref{Krr}) and (\ref{hamilton}) are
integrated to find ${K^r}_r$ and $R$.  
And then eqn (\ref{lapse}) is solved for the lapse $\alpha$
(using a tridiagonal method and the fact that 
${\partial _r}\alpha =0$ at the origin and $\alpha \to 1$ at infinity).
Then eqn (\ref{shift})
is integrated to find the shift $\beta ^r$.  Finally, the quantities $\phi , \, S$ and $P$ are
evolved to the next time step using eqns (\ref{dtphi}), (\ref{dtS}) and (\ref{dtP}) respectively.  The 
evolution is done using the iterated Crank-Nicholson method, and all spatial derivatives are found using standard centered 
differences.  We use units where $\kappa =1$.  

As the evolution proceeds we can check for black hole formation by looking for the presence of a marginally outer
trapped surface.  In spherical symmetry, such a surface is given by the condition ${\nabla ^a}R{\nabla _a}R=0$ which
in our coordinate system is equivalent to 
\begin{equation}
{\partial _r} R + {\textstyle {\frac 1 2}} R {{K^r}_r} = 0
\end{equation}
 
\section{Results}

\begin{figure}
\includegraphics{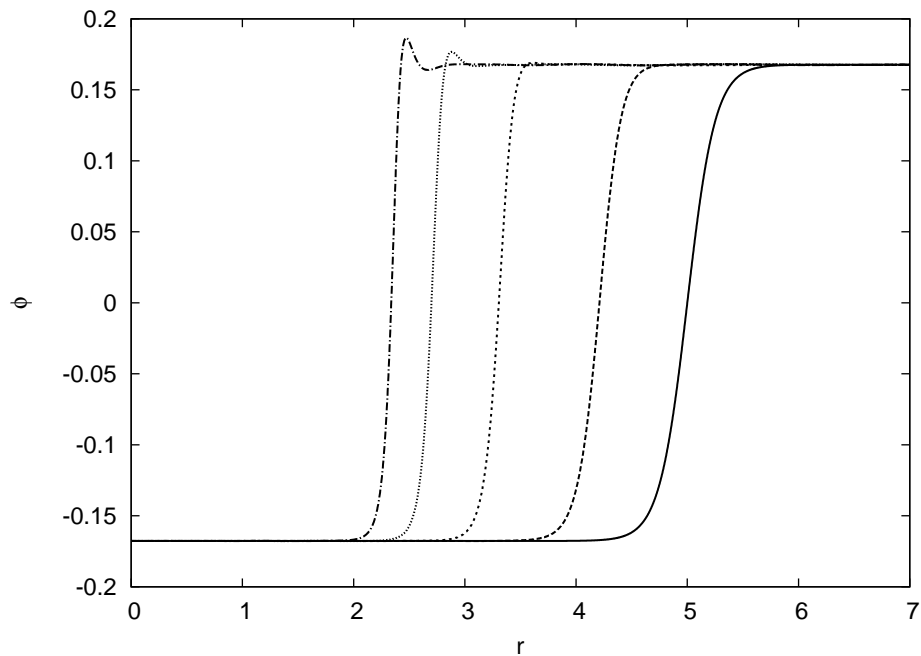}
\caption{\label{fig1} The scalar field $\phi$ at times 0, 2,  4, 6 and 8}
\end{figure}

The simulations are run with $n+1$ points evenly spaced between $r=0$ and a maximum value $r_{\rm max}$.  
We choose $r_{\rm max} =20$.  
We would like to know how the initial scalar field profile changes under evolution.  Fig \ref{fig1} shows the result of 
a simulation with ${r_0}=5, \, \epsilon = 0.25$ and $\sigma = 0.15$.  This simulation was run with $n=9600$ and results are
shown for the scalar field $\phi$ at times 0, 2, 4, 6, and 8.  As the evolution proceeds, the scalar field profile moves inward.  Furthermore, the profile becomes steeper, and departures from the simple $\tanh$ form become 
more pronounced.  A marginally outer trapped surface forms at $t=5.09$.  

\begin{figure}
\includegraphics{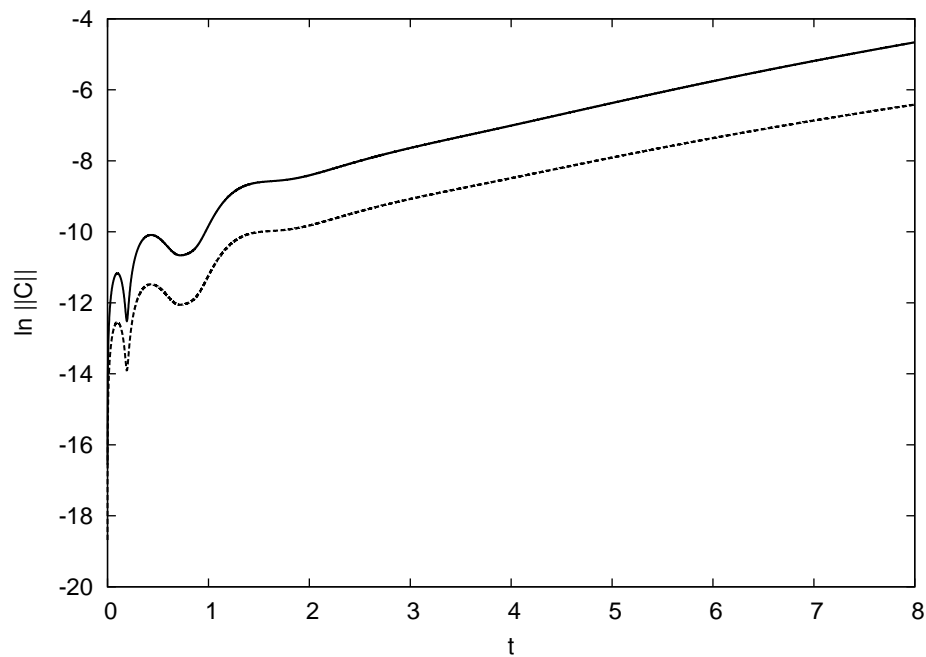}
\caption{\label{fig2} $\ln ||{\cal C}||$ vs. $t$ for $n=9600$ (solid line) and 
$n=19200$ (dashed line).  The results demonstrate second order convergence.}
\end{figure}

In order to be sure that these results are not numerical artifacts, we need to know that the code is convergent and that these 
results are within the convergent regime.  Note that eqn. (\ref{dtR}) is not used in the evolution, but should nonetheless 
be satisfied to within numerical error due to finite differenceing.  Thus this equation provides a check on the performance 
of the simulation.  More precisely, define the constraint quantity $\cal C$ by
\begin{equation}
{\cal C} = {\frac \alpha 2 }{{K^r}_r} + {R^{-1}}({\beta ^r}{\partial _r}R - {\partial _t}R)
\end{equation}
and let $||{\cal C}||$ be the $L_2$ norm of $\cal C$.  Then in an exact solution $\cal C$ should vanish, while in a
numerical treatment $\cal C$ should converge to zero in the limit of zero step size.  Fig \ref{fig2} shows 
$\ln ||{\cal C}||$ as a function of time.  Here the parameters are as in the previous simulation, except that one simulation
is run with $n=9600$ and one with $n=19200$.  The results demonstrate second order convergence: that is, halving the step size 
reduces $\cal C$ by a factor of 4.    

In the thin shell formalism one is mostly concerned with the motion of the wall.  The motion 
of a spherical wall is described
by giving its area radius as a function of proper time.  That is, 
\begin{equation}
R = {R_0}(\tau)
\end{equation}
Using the results of \cite{ipser,israel} one can show that the equation of motion of the thin shell is 
\begin{equation}
{{\ddot R}_0} = {\textstyle {\frac 3 4}} \kappa \sigma {{(1 + {{\dot R}^2 _0})}^{1/2}} - 2 {R_0 ^{-1}}
(1 + {{\dot R}^2 _0})
\label{thinmotion}
\end{equation}
  
\begin{figure}
\includegraphics{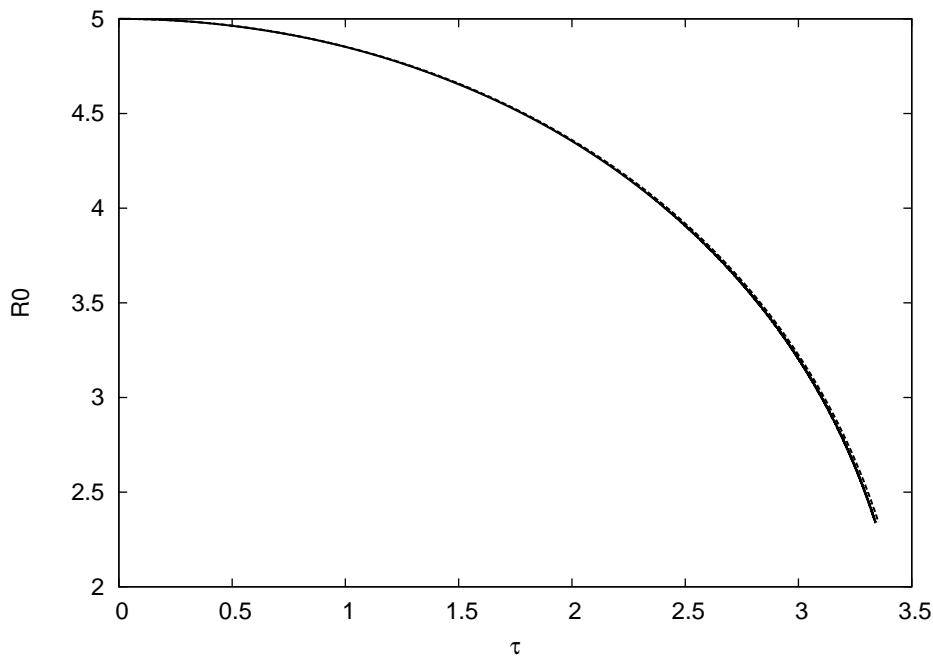}
\caption{\label{motionfig} $R_0$ vs. $\tau$ for the simulation (solid line) and 
the thin wall equations of motion (dashed line).}
\end{figure}

We would like to know how well the motion given by solving eqn. (\ref{thinmotion}) models the behavior of the 
scalar field domain wall.  Here we can do a direct comparison: at any given time, the position of the wall will be taken to be 
the place where $\phi =0 $ and the proper time $\tau$ will be that of an observer who is always at the position of the
wall.  At each time step of the simulation, we can find the position of the wall, so it remains to evaluate $\tau$.  
Let $u^a$ be the four-velocity of the observer who remains at the position of the wall.  Then it follows that $u^a$ is a 
unit timelike vector for which ${u^a}{\nabla _a}\phi = 0$.  It then follows 
using eqns (\ref{metricform}) and (\ref{PSdef}) that 
\begin{equation}
{u^a} = {A^{-1}} \left [ {n^a} - {\frac P S} {{\left ( {\frac \partial {\partial r}} \right ) }^a} \right ]
\label{wallvelocity}
\end{equation}
where the quantity $A$ is defined by 
\begin{equation}
A = {\sqrt {1 - {\frac {P^2} {S^2}}}}
\end{equation}
evaluated at the position of the wall.  The relation between the normal vector and the time coordinate is 
\begin{equation}
{n_a} = - \alpha {\nabla _a}t
\end{equation}
so using eqn. (\ref{wallvelocity}) we have  
\begin{equation}
{\frac {dt} {d\tau}} = {u^a}{\nabla _a} t = - {\alpha ^{-1}} {u^a}{n_a} = {\alpha ^{-1}}{A^{-1}}
\end{equation}
and therefore we find
\begin{equation}
d \tau = \alpha A d t
\label{taumotion}
\end{equation}
Choosing $\tau =0$ at the begining of the simulation, we then integrate eqn. (\ref{taumotion}) to find $\tau$ at 
all times of the simulation.  Since we also have $R_0$ at all points of the simulation, we can produce the thick 
wall ${R_0}(\tau)$ for comparison with the ${R_0}(\tau)$ of the thin wall.  Fig. (\ref{motionfig}) shows such a
comparison.  Here the parameters of the wall are the same as in the previous simulations, with the solid line
being that of the simulation and the dashed line the solution of eqn. (\ref{thinmotion}).  

We now consider a comparison of the stress-energy of the simulation to that of the thin shell approximation.  
From eqns (\ref{stressenergy}) and (\ref{PSdef}) it follows that the trace of the stress-energy tensor is 
\begin{equation}
T = {P^2} - {S^2} - 4 V
\label{Tthick}
\end{equation}
However, in the treatment of \cite{ruthandme1} to lowest order in thickness of the wall we have
\begin{equation}
T =  {\frac {- 9 \sigma} {4 \epsilon {\cosh ^4} (z/\epsilon)}}
\label{Tthin}
\end{equation}
where $z$ is geodesic distance from the center of the wall.  Calculating geodesic distance from a simulation
can be complicated; but luckily because the expression for $T$ in eqn. (\ref{Tthin}) falls off so rapidly, 
we only need $z$ in the vicinity of the wall's center and thus can approximate $z$ using a
Taylor series.  In particular, note that at the center of the wall ${\nabla _a} z$ must be a unit vector 
orthogonal to $u^a$.  It then follows that 
\begin{equation}
{\nabla^a} z = {A^{-1}} \left [  {{\left ( {\frac \partial {\partial r}} \right ) }^a}  - {\frac P S}
{n^a} \right ]
\label{gradz}
\end{equation}
Taking the inner product of eqn. (\ref{gradz}) with ${(\partial /\partial r)}^a$ we find that at the center of the wall
\begin{equation}
{\frac {\partial z} {\partial r}} = {A^{-1}}
\end{equation}  
and we therefore find that near the wall $z$ is well approximated by
\begin{equation}
z = (r-{R_0})/A
\label{zapprox}
\end{equation}
Thus the thin wall expression for $T$ is given by eqn. (\ref{Tthin}) with $z$ given by eqn. (\ref{zapprox}).

\begin{figure}
\includegraphics{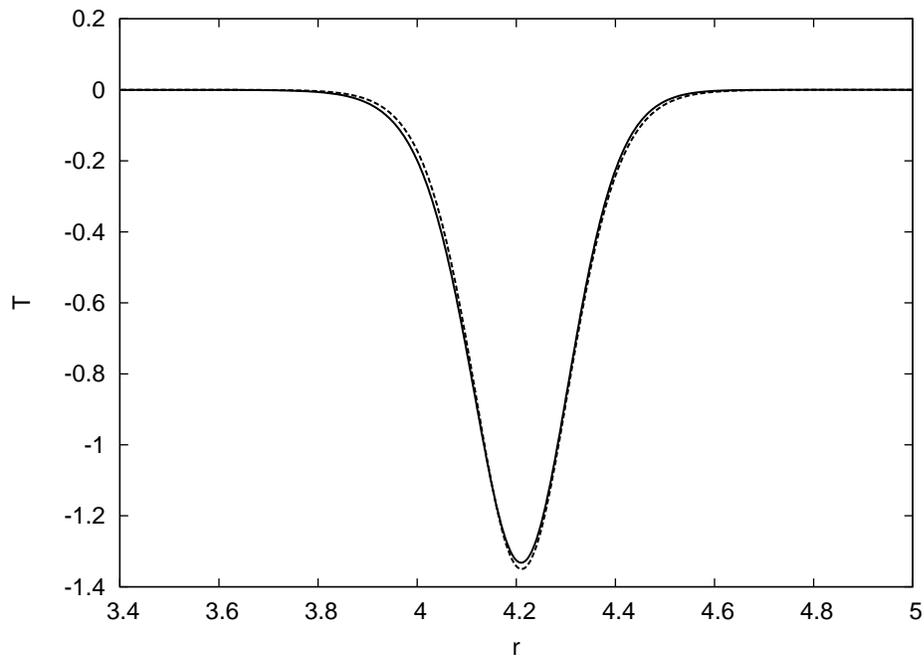}
\caption{\label{energyfig1} $T$ vs. $r$ at $t=2$ for the simulation (solid line) and 
the thin wall approximation (dashed line).}
\end{figure}

\begin{figure}
\includegraphics{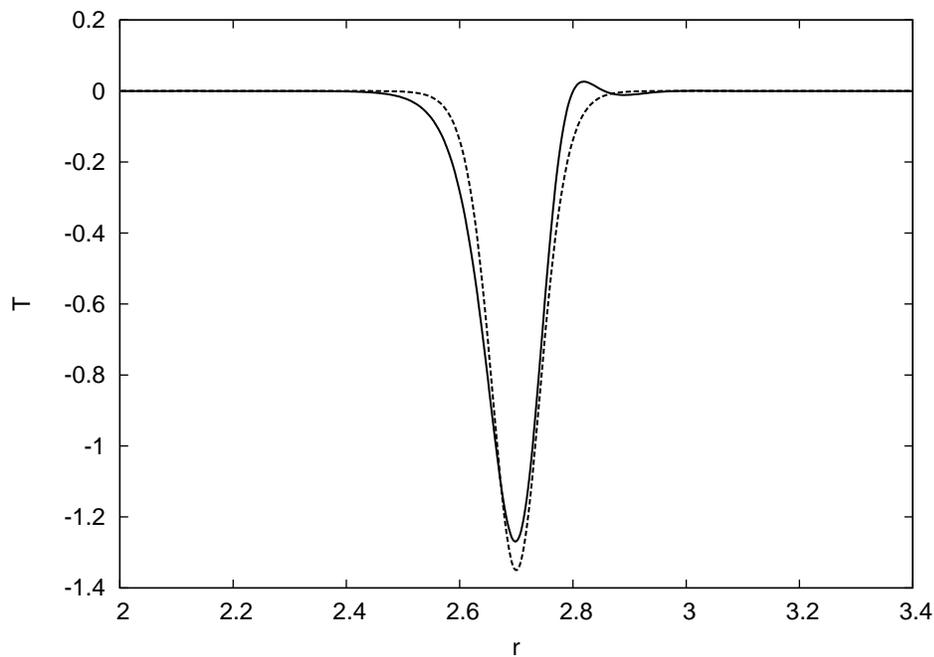}
\caption{\label{energyfig2} $T$ vs. $r$ at $t=6$ for the simulation (solid line) and 
the thin wall approximation (dashed line).}
\end{figure}

Such a comparison is given in Fig (\ref{energyfig1}) for $t=2$ and in Fig (\ref{energyfig2}) for $t=6$.  
The parameters of the wall are the same as in the previous simulations.
Here the trace of the stress-energy $T$ is plotted
for the simulation (solid line representing the expression of eqn. (\ref{Tthick})) and for the thin shell approximation
(dashed line representing the expression of eqn. (\ref{Tthin}).  Note that the simulation results are well approximated by
the thin shell expression, though the approximation becomes less good as the evolution proceeds.

\section{Conclusions}

Our simulations indicate that the thin wall approximation is an excellent 
approximation for thick wall gravitational collapse.  Perhaps more importantly, we have developed a
robust numerical method for thick wall collapse that could be used on other projects.  In particular,
we plan to simulate the collapse of a charged thick domain wall.  Thin charged walls collapse to 
charged black holes and can even be used to form an extreme (charge equals mass) black hole.\cite{israel2}  In 
contrast, simulations of the collapse of a free charged scalar field yield black holes that are always far from
extreme.\cite{piran}  It will be interesting to see whether an extreme black hole can be formed by the collapse
of a charged thick domain wall.

\ack
This work was supported by NSF grant PHY-0855532 to Oakland University.


\section*{References}
\begin{thebibliography}{}

\bibitem{ipser}
Ipser J and Sikivie P 1984 {\it Phys. Rev.} {\bf D30} 712

\bibitem{israel}
Israel W 1966 {\it Nuovo Cimento} {\bf 44B} 1 

\bibitem{ruthandme1}
Garfinke D and Gregory R 1990 {\it Phys. Rev.} {\bf D41} 1889

\bibitem{ruthandme2}
Gregory R, Haws D, and Garfinkle D 1990 {\it Phys. Rev.} {\bf D42} 343

\bibitem{widrow}
Widrow L 1989 {\it Phys. Rev.} {\bf D40} 1002

\bibitem{spergel}
Press W, Ryden B, and Spergel D 1989 {\it Astrophys. J.} {\bf 347} 590

\bibitem{matt}
Choptuik M, {\it Phys. Rev. Lett.} 1993 {\bf 70} 9

\bibitem{ratinryome}
Akhoury R, Garfinkle D, and Saotome R 2011 {\it JHEP} {\bf 1104} 096

\bibitem{israel2}
de la Cruz V and Israel W 1967 {\it Il Nuovo Cimento} {\bf 51} 744

\bibitem{piran}
Hod S and Piran T 1997 {\it Phys. Rev. D} {\bf 55} 3485

\end{thebibliography}
\end{document}